# The impact of perceived recognition by physics instructors on women's self-efficacy and interest

Yangqiuting Li and Chandralekha Singh

*Department of Physics and Astronomy University of Pittsburgh, Pittsburgh PA 15260*

**Abstract.** Students' self-efficacy, interest, and perceived recognition from others in a given field have been shown to be very important for the development of their identity in that field, which is a critical predictor of students' major and career decisions. Prior research suggests that students' self-efficacy and interest play an important role in their engagement, performance, and persistence in STEM fields. However, very little has been investigated about the role of perceived recognition and validation by instructors on students' self-efficacy and interest. Moreover, prior quantitative studies show that women often report a lower level of physics perceived recognition, self-efficacy and interest. In this study, we analyzed data from individual interviews with 38 female students to investigate their learning experiences in physics courses in order to obtain a qualitative understanding of the factors that shape their self-efficacy and interest. We find that female students' negative and positive perceived recognition from instructors and teaching assistants (TAs) greatly influenced their self-efficacy and interest and even impacted their desire to persist in STEM majors. We categorize different types of perceived recognition that women reported in our interviews and how they influenced them. For example, many women reported that they felt belittled for their questions or efforts in physics courses, which often negatively influenced their self-efficacy. These findings can help physics educators develop better ways to interact with students in order to provide positive recognition and validation, such as acknowledging students' efforts and questions, expressing faith in students' ability to excel, and being careful not to give unintended messages to students. Our research also suggests that it is important for instructors/TAs to note that it is not their intentions that matter but the impact they are having on their students.

## I. INTRODUCTION

In the disciplines of science, technology, engineering, and mathematics (STEM), there have been efforts to enhance the participation and advancement of marginalized groups such as women [1-15]. Prior research suggests that individuals' course enrollment, degree attainment and achievement in STEM can be influenced by domain specific motivational factors such as self-efficacy, interest and identity [2,7,12,16-20]. For students from marginalized groups, these motivational variables might be impacted, e.g., by negative societal stereotypes and biases about who belongs and can excel in STEM as well as lack of role models and encouragement from others, which can lead to withdrawal from STEM courses, majors or careers [21-26]. Hence, investigating motivational factors is critical to understanding and addressing diversity, equity, and inclusion issues in STEM disciplines.

For explaining participation in STEM careers, identity has been argued to be a particularly important motivational construct [1,2,5,27]. Students' identity in an academic domain, such as physics, relates to their views about whether they see themselves as a "physics person" [1,5], and it has been shown to influence students' career decisions as well as short- and long-term academic goals [2,5,27]. Prior studies suggest that students' physics identity includes three interrelated dimensions: perceived recognition by others as a physics person, self-efficacy with respect to physics, and interest in physics [5,28,29]. Prior quantitative studies using structural equation modeling show that these three dimensions (perceived recognition, self-efficacy and interest) are important predictors of the extent to which students see themselves as a physics person (which is a holistic measure of students' overall physics identity) [27,30-32].

Perceived recognition in a domain, such as physics, refers to students' perception about whether other people see them as a physics person [33]. Prior quantitative studies show that perceived recognition is the strongest predictor of students' overall physics identity as compared to self-efficacy and interest [27,28]. Moreover, perceived recognition has also been shown to predict students' course grades in introductory physics courses [34,35].

Self-efficacy, which is defined as students' beliefs in their capability to succeed in a certain situation, task, or particular domain [28,36,37], has been shown to influence students' engagement and performance in a given domain [16,18]. Students with high self-efficacy in a domain often enroll in more challenging courses in that domain than those with low self-efficacy because they perceive difficult tasks as challenges rather than threats [17].

Interest is defined by positive emotions accompanied by curiosity and engagement in particular content [38]. Interest has also been shown to influence students' learning [16,19,38]. For example, one study showed that making science courses more relevant to students' lives and transforming curricula to promote interest in learning can improve students' achievement [39].

Prior quantitative studies have shown that female students often report a lower level of self-efficacy and interest than their male peers in physics [40-43]. For example, a prior study showed that female students with an A grade had similar physics self-efficacy as male students with a C grade by the end of a two-semester introductory calculus-based physics course (whereas these women with A grades had the same self-efficacy as men with B grades at the end of the first-semester physics course) [43]. In other words, the gender difference in self-efficacy increased over time. This study is consistent with another study showing that female students are more likely to drop STEM majors with significantly higher grade point averages than their male peers [44]. In order to shed light on these types of quantitative findings, qualitative studies probing the experiences of female students in physics courses can be invaluable. These qualitative studies can also shine light on whether women have differential experiences compared to men in physics courses, how these experiences shape women's physics self-efficacy and interest, and what instructors and TAs could do to create an equitable and inclusive learning environment and better support female students in these courses.

Therefore, in this study, we conducted individual interviews with 38 women in both introductory and advanced physics courses at a large research university in the US to investigate their experiences and how these experiences shape their self-efficacy and interest. In particular, we focused on the relationships between perceived recognition and the other two dimensions of physics identity, i.e., self-efficacy and interest. Our results can be valuable for formulating guidelines for creating an inclusive learning environment in which all students can excel.

## II. THEORETICAL FRAMEWORK

As noted earlier, self-efficacy, interest, and perceived recognition from others are three interrelated dimensions of physics identity. This physics identity framework is adapted from prior studies focusing on domain-specific identity [1,5,45]. In Carlone and Johnson's science identity framework [1], students' science identity includes three interrelated dimensions: competence (belief in one's competence), performance (belief in ability to perform), and recognition (recognition of self and by others as a "science person"). Hazari et al. adapted this model to physics and added interest to this model [5]. In addition, Hazari et al. developed quantitative measures for these constructs and found that competence and performance factored into a single construct [5]. Moreover, they separated recognition of self and by others and used a single item ("I see myself as a physics person") as a holistic measure of students' overall physics identity [5]. In Hazari et al.'s later quantitative studies using structural equation modeling, they found that students' overall physics identity was predicted by their interest, competence/performance beliefs, and perceived recognition from other people [27,30-32]. The physics competence/performance beliefs is peoples' beliefs about their ability to understand and perform physics [5], which is very similar to the definition of self-efficacy. Kalender et al. adapted Hazari et al.'s physics identity framework and used self-efficacy in place of performance/competence beliefs in their model [45].

Prior studies have shown that female students did not feel that they were recognized appropriately in many science domains such as physics [46,47]. For example, a report from the National Science Foundation [48] indicated that elementary and high school boys and girls interested in science felt that they were treated differently by parents, teachers and friends with regard to encouragement and engagement in science. While boys reported receiving admiration and encouragement, girls reported interactions and responses from others that were often characterized as ambivalent, with lack of encouragement, or suggestions that their goals were inappropriate [48]. Prior studies show that similar biases and stereotypes also impact female students in the university context [28,49]. For example, one study showed that science faculty members in biological and physical sciences exhibited biases against female students and rated male students as more competent, and they were more likely to hire and mentor them and pay them more even though only the names were different in the hypothetical information faculty members were provided about the student [49].

The experiences of not being recognized as a science person have the potential to not only impact students' science identity directly, they may also influence students' self-efficacy and interest in science, which are the other two dimensions of science identity. Prior studies indeed suggest that the three dimensions (self-efficacy, interest, and perceived recognition) of physics identity are not independent constructs [16,28,42,50-52]. For example, according to Eccles's Expectancy-Value Theory (EVT) [52,53], interest is often connected with self-efficacy. In particular, expectancy refers to students' belief in their ability to succeed in a given task [52] (which is closely related to self-efficacy) and value refers to the subjective task value for students [52] (which is closely related to interest). The EVT suggests that expectancy and value interact to affect students' engagement, academic achievement, and persistence in a field. Moreover, prior studies have also shown that interest in a domain can be affected by self-efficacy [50,51]. Furthermore, although prior quantitative studies have shown that perceived recognition is also correlated with self-efficacy and interest [28,42], to our knowledge, no prior studies have qualitatively investigated the relationships between perceived recognition, especially from instructors and teaching assistants (TAs), and the other two dimensions of physics identity, i.e., self-efficacy and interest.

Therefore, the theoretical framework of our qualitative study discussed here focuses on unpacking the relationship between female students' perceived recognition from physics instructors/TAs and their physics self-efficacy and interest. We centered the experiences of female students in physics courses from introductory to advanced levels in order to investigate how their interactions with instructors/TAs may be affecting them so that we can contemplate how to support them and improve the physics learning environment. In particular, since instructors and TAs are the ones who have the power to empower students, investigations focusing on female students' perceived recognition, especially from physics instructors and TAs, are valuable in developing guidance on how instructors/TAs can create an equitable and inclusive physics learning environment in which students feel supported and recognized positively.

Inspired by our framework, we conducted a qualitative study involving individual interviews with female students to investigate their experiences in physics courses. In particular, we focused on female students' perceived recognition from instructors and TAs and how it might shape the other two dimensions of physics identity, i.e., self-efficacy and interest. In the interviews, we also asked students questions about their academic background, overall college experience, and interaction with peers to get a better understanding of their experiences.

## III.     RESEARCH QUESTIONS

Consistent with our theoretical framework, we seek to address the following research questions:

**RQ1.** What are female students' perceptions of different types of recognition from instructors/TAs in physics courses?

**RQ2.** How do female students' perceived recognitions from instructors/TAs shape their self-efficacy and interest?

## IV.     METHODOLOGY

### A.   Participants

We conducted semi-structured, empathetic interviews with 38 female students in physics courses and majoring in Physics and Astronomy, Engineering/Computer science, or Chemistry (includes those on a bio track) at a large research university in the US. Although 22 of these women were White, 12 were Asian, 3 were Hispanic and 1 was Black, our focus here is on their experiences as women in physics. Our choice of focusing on women in physics courses is inspired by the framework of standpoint theory [54,55], which emphasizes that one should center the experiences of people from traditionally marginalized groups since they have experienced inequities and will be able to articulate how those inequities manifest in their everyday experiences and point to the issues within the system that must be fixed. The self-reported year and discipline are shown in Table I. We sent the interview advertisement to the department's undergraduate administrative assistant to share it with all of the undergraduate students in the physics department, and we also sent the advertisement to instructors of different physics courses and asked them to share it with their students. Each student received a $25 gift card for participating in an hour-long interview. Roughly half of these students were interviewed before the COVID-19 pandemic and half of them during the pandemic. The interviews followed a semi-structured think aloud protocol with interview questions that were agreed upon by the researchers prior to the interviews [56]. We call these interviews empathetic interviews because the goal of the interviews was to understand the experiences of female students in physics courses in order to improve equity and inclusion. In particular, before interviews, we communicated with students that the broader goal of the research is to make the physics learning environments equitable and inclusive. All participants agreed to be audio-recorded and quoted in academic publications. Students also had the opportunity to ask questions about the research before and after the interviews.

**Table I.** Self-reported year and discipline of the female students (N=38) interviewed.

| Year | |
|---|---|
| First year | 20 |
| Second year | 8 |
| Third year | 3 |
| Fourth year | 6 |
| First year graduate student talking about undergraduate experience | 1 |
| Discipline | |
| Physics and Astronomy | 19 |
| Engineering/Computer science | 17 |
| Chemistry | 2 |

### B. Semi-structured interviews

To investigate female students' learning experience in physics courses and their perception of the learning environment, we assembled and refined a list of potential interview questions via an iterative process between the researchers. These included questions about the student's background (e.g., how they got interested in STEM, early experiences in K-12 including high school experience); overall college experience so far; experience in physics courses (such as their interaction with instructors, TAs and peers) both inside and outside of the classroom; and perception of their learning in physics courses (including any challenges in learning, thoughts on underrepresentation of women and how to improve the physics learning environments). In particular, inspired by our theoretical framework [5,28], we focused on female students' perceived recognition from instructors and TAs and investigated how the perceived recognition from instructors and TAs impacted the other two dimensions of physics identity, i.e., self-efficacy and interest. Therefore, many questions in the interview are used to elicit students' perceived recognition from instructors and TAs. Examples of these types of questions are shown in Table II. As shown in Table II, many interview questions are neutral, for example, "Do you feel supported by your instructors and TAs/UTAs in your physics course(s)?" and "How do you think your experiences in your physics classes might have affected your identity as a physics person?". Even though the questions such as "Have you witnessed any barriers to success that students in physics at this institution have experienced because of their gender?" might prime the students towards negative reflection, we also asked many questions that help student reflect on their positive experiences. For example, "Have you had any role models in your classes in terms of professors/TAs etc.? How has this impacted you?" and "Are there any programs or services offered by the department or your professors in physics that you have found useful?" The goal of asking these different types of questions is to get a better understanding of different aspects of students' perceived recognition. Moreover, since prior studies have shown that women often report lower perceived recognition, self-efficacy and interest than male students in physics courses [41-43], it is useful to explicitly ask female students about their thoughts on how their gender may influence their experiences in physics courses.

In our study, we made the semi-structured hour-long individual interviews empathetic to give students the opportunity to express themselves freely, dig deeply on critical issues of equity and inclusion in physics courses, and make sure they felt comfortable expressing themselves. We asked students to think aloud as they answered the questions, and we did not disturb them when they thought aloud in order to not disrupt their thought processes and asked them for clarifications of the points they may not have made only after they had finished their thoughts. Most participants required little prompting and were very keen to share their thoughts with us openly. All interviews were audio-recorded and transcribed.

**Table II.** Examples of the interview questions that can elicit students' perceived recognition from instructors and TAs.

| Example questions |
| --- |
| How do you think your experiences in your physics classes might have affected your identity as a physics person? |
| Do you feel respected as a physics person/someone who can excel in physics courses by your peers, TAs, and instructors? |
| Do you feel supported by your instructors and TAs/UTAs in your physics course(s)? |
| Do you feel that your contributions are valued by your peers, TAs, and instructors? |
| Have you had any role models in your classes in terms of professors/TAs etc.? How has this impacted you? |
| Are there any programs or services offered by the department or your professors in physics that you have found useful? |
| Do you think your experiences in physics have been different because of your gender? If so, how? |
| Do you think your gender has had an impact on your success as a physics student/a student in physics courses? |
| Have you witnessed any barriers to success that students in physics at this institution have experienced because of their gender? |
| Do you think that it is more difficult to succeed in physics/physics courses as a woman? Why/why not? |

### C. Data analysis

We coded the interviews using hybrid coding methods that involved both deductive and inductive coding [57]. Initially, deductive coding was used based on the interview protocols, but after reading through the interviews, we incorporated inductive coding to encompass different aspects of the interviewed women's experiences. The first author coded the interviews using Nvivo and both authors discussed and converged on the codes developed based on the student interviews. The codes are inspired by our theoretical framework and interviews. As noted, our framework focuses on female students' perceived recognitions from instructors or TAs and the relationships between perceived recognition and the other two dimensions of physics identity, i.e., self-efficacy and interest. Both authors discussed how the codes could be combined to form larger themes. In all, 43 codes resulted in 5 broad analytic themes (Perceived recognition from instructors and TAs, High school and other pre-college experiences, Interactions with peers, Research experiences, and Suggestions to improve physics learning environments). In this paper, consistent with our framework, we focus on the first theme, perceived recognition from instructors and TAs.

## V. RESULTS

In our interviews, 26 (68%) of the interviewed female students reported negative perceived recognition or lack of positive recognition from their physics instructors or TAs, while only 4 students (11%) reported positive perceived recognition. The other students did not provide answers to questions related to perceived recognition that were clearly categorized as positive or negative by researchers. We divide the interview findings into three broad sections: (i) Negative perceived recognition or lack of positive recognition from instructors/TAs; (ii) Impact of negative perceived recognition or lack of positive recognition from instructors/TAs on students' self-efficacy and interest; (iii) Positive recognition from instructors/TAs and its influence on students' self-efficacy.

### A. Negative perceived recognition or lack of positive recognition from instructors/TAs

In our interviews, 68% of the interviewed female students reported negative perceived recognition or lack of positive recognition from their physics instructors/TAs. The negative perceived recognition or lack of positive recognition by female students from instructors/TAs can be categorized into four subthemes: (1) Feeling belittled for questions or efforts; (2) Feeling marginalized due to differential gender dynamics; (3) Feeling that the physics learning environment is unsafe; (4) Feeling negatively recognized about their abilities and potential. As shown in Table III, each subtheme includes two codes, and each code is followed by a definition and an illustrative example from the participants in this study.

**Table III**. Subthemes for negative perceived recognition or lack of positive recognition from instructors/TAs. The percentages in parentheses represent the proportions of the interviewed women whose experiences were coded under each subtheme. In total, 68% of the interviewed female students reported perceived recognitions that were coded under at least one of the following subthemes.

| Subtheme | Code | Definition | Example from participant data |
| --- | --- | --- | --- |
| Feeling belittled for questions or efforts (17/38 = 45%) | Belittling students' questions or struggles | Expressing that students' questions or some concepts that students are struggling with are easy, obvious or trivial | *"…I thought this was so easy… I'm disappointed you guys didn't get this."* |
| | Belittling students' efforts | Negative recognition or lack of positive recognition for students' efforts, improvement, and achievement | *"Once he finally comes over to help me, he doesn't actually acknowledge any of the work that I've done…he doesn't try to find a solution from what I started, he just does it his own way."* |
| Feeling marginalized due to differential gender dynamics (14/38=37%) | Differential treatment of female and male students | Responding to or treating men and women differently when interacting with students | *"I've interacted with, very sadly, a TA who has sort of brushed off my question and then answered a guy's question in my group…"* |
| | Letting men dominate the class | Letting men dominate the class so that the classroom dynamics dominated by men make women feel marginalized | *"They had men ask and answer questions and everyone else was sitting quietly"* *"Once I asked a question in office hour, he asked a guy to do it for me on the board."* |
| Feeling that the physics learning environment is unsafe (11/38=29%) | Condescending or intimidating behavior | Displaying a feeling of patronizing superiority or causing feelings of fear in students about asking questions | *"My professor was pretty condescending, just not really willing to help… I was personally very terrified of him. I never wanted to go to him with anything."* |
| | Cold calling | Cold calling students randomly to answer questions or show work in front of their peers | *"He would ask you to write stuff on the board in front of the whole group. It was very intimidating."* |
| Feeling negatively recognized about their abilities and potential (6/38=16%) | Underestimating students' abilities | Doubting some students' ability to do well in a task or having low expectations of some students | *"He was like leaning over my shoulder and like telling me one key at a time what to type instead of just trusting me to be able to spell the word."* |
| | Fixed mindset about students' potential | Emphasizing brilliance rather than efforts | *"He definitely thinks that some people cannot do physics."* |

### *1. Feeling belittled for questions or efforts*

In our analysis, 45% of the interviewed female students reported experiences that were coded under the subtheme *feeling belittled for questions or efforts*, which includes two codes: *belittling students' questions or struggles* and *belittling students' efforts*.

### *(a) Belittling students' questions or struggles*

This code includes instances in which physics instructors and/or TAs express that students' questions or some concepts that students are struggling with are easy, obvious or trivial. Often these comments were by instructors and TAs explicitly using the words "easy", "trivial", and "obvious", which made these female students feel that their questions were not good and they did not want to ask more questions. Hailey shared an experience from her Electricity and Magnetism class:

> *Hailey: He [the instructor] was also the kind of professor that would say like, "oh, all this information is trivial, so let me just like [go over it] really fast and assume that you know it" ... So I just felt like I didn't want to go and like ask a stupid question to him.*

Mary shared a similar experience in her physics class:

> *Mary: My physics professor said, like, that, things were obvious a lot. And according to my TA, who's worked closely with him, he [the instructor] says that to make it [the question] not seem as overwhelming. But for me, at least, the effect was much more like, if you don't know this, you're dumb. So that made it difficult to like, ask questions in class.*

Thus, Hailey and Mary note that the comments from their physics instructors make them hesitate to ask questions to their instructors because they are worried that their questions will look "stupid". Mary also mentioned that even though her TA thinks that the instructor may want to make students feel at ease and encourage them by saying that the problems are easy or trivial, these comments have a negative impact on her and make her feel "dumb" rather than encouraged.

Sometimes, instructors may not explicitly use words such as "easy", "trivial", "obvious", etc. directly; however, the interviewed students reported that the way instructors communicate with students can also make them feel nervous about not knowing and asking questions. Fem shared her experience of interacting with her physics instructors:

> *Fem: Sometimes when they [the physics instructors] are explaining something, they kind of make you feel dumb. It's never their intention, it [is] just clearly in their tone [and] in the way that they talk, it's like you should know this…I don't like going to my teachers, because sometimes they make me feel dumb.*

Elaine recalled a similar experience in a physics instructor's office hours:

> *Elaine: I had some bad experiences [in office hours], where like, I would ask a question, and then I'd just feel the way he would explain it would make me feel like really dumb ... because I had like a few and, what I think are embarrassing moments. I stopped going to office hours, to my professor's office hours in fall semester.*

As we can see from both Fem and Elaine's experiences, even without explicitly saying something is "easy" or "trivial", the tone or body language of instructors can also convey similar message that can discourage students. In addition, instructors may not even realize it when they are conveying this kind of message to students. Fem later noted that "*It's never their intention*";

however, it still negatively influenced her. It is important for instructors to understand that it is really the effects of instructors on students that matter rather than instructors' intentions.

In addition to body language, when instructors respond to students' questions quickly and carelessly or they show lack of interest in students' questions, this can also convey the message that students' questions are not good, worth asking or important. Hailey recalled that in office hours, her physics instructor seemed to be more interested in hard physics questions (usually asked by male students), which made her nervous about asking homework questions.

> *Hailey: Even though he's like, "Oh yeah, come ask me questions", he didn't really seem like he was interested in helping me… he seemed like he was interested in answering what he considered to be good questions about physics whereas like, I just wanted to know how to do homework, you know?*

Shreya narrated that the quick and careless response from her physics instructor was very discouraging for her.

> *Shreya: When I'd asked him about like specific things, he'd just be like, "Oh, I think you should just review this topic" …I think it definitely deterred me from asking more questions.*

As we can see from the experiences of the interviewed female students in physics courses, the discouraging and belittling comments from instructors made students feel less comfortable or outright uncomfortable to ask questions because they worried that their questions were not good enough and felt nervous about showing that they did not know something. Moreover, many female students reported that these discouraging behaviors of physics instructors are contagious. For example, Samantha mentioned that she noticed that after physics professors used "easy", "trivial", and "obvious" in their lectures, her male peers started to pick up these cues and used these terms.

> *Samantha: "Oh, this is trivial, and you should know this right?" I did not encounter that phrase during high school. And then when I came to college, I heard professors start to use that. And then shortly after I heard the professors use that, I started hearing my [male] peers say that, and I feel like it's something that's almost like also taught along with the curriculum, [it] is like, this is how physics culture is.*

This experience of Samantha is consistent with our interviews with other female students, many of whom reported being also talked down and disparaged by their male peers during physics learning.

*(b) Belittle students' efforts*

This code refers to situations in which students described receiving negative recognition or lack of positive recognition for their efforts, improvement, and achievement. For example, Suzie shared the following example of her instructor giving negative recognition to students routinely:

*Suzie: The instructor usually begins the classroom by saying "You guys aren't serious about this class because you're not doing well on the work", and I'm just like you don't have to tell me that, you don't have to start every class with that, I don't want to hear.*

Suzie mentioned that these comments discouraged her and her peers from working hard in this course because their work was never recognized by the instructor. In addition to negative recognition, lack of positive recognition can also discourage students. For example, Maya noted that even small recognition or acknowledgement could encourage her; however, she never gets any when asking questions to her college physics instructors and TAs:

*Maya: [When asking questions to physics instructors or TAs], I usually clearly show that I'm reading the material and I understand what's going on, but there's a certain concept that I don't understand. However, I've never been told directly like, "Oh, you're doing a good job", or like, "No, you got it, you're just making a simple mistake" or something like that. So I guess I never really had the reinforcement, from a professor or TA of like, "You're being too hard on yourself".*

Lack of positive recognition or acknowledgement can make students worry about whether their work is good enough even though they might actually have done well, which may impact students' self-assessment and self-efficacy [58]. This is especially true for students who are underrepresented in physics, a field with strong stereotypes about who belongs and can excel in. Lack of positive recognition has negatively impacted the entry and retention in physics related disciplines for decades.

### 2. Feeling marginalized due to differential gender dynamics

In our analysis, 37% of the interviewed female students reported experiences that were coded under the subtheme *feeling marginalized due to differential gender dynamics*, which includes two codes: *differential treatment of female and male students* and *letting men dominate the class.*

#### (a) Differential treatment of female and male students

This code refers to situations in which instructors and/or TAs respond to or treat women and men differently when interacting with students. For example, Amy shared her experience in a college physics recitation:

*Amy: There have also been times, where I've interacted with, very sadly, a TA who has sort of brushed off my question and then answered a guy's question in my group or came into the room and only greeted the guys in my group and not greeted me.*

Amy further mentioned that even though she thinks this experience is not necessarily aggressive or outright offensive, it makes her feel that there is a difference in the treatment of men and women in this field. In our interviews, we found that the differential treatment of women and men also happens in other STEM courses and sometimes in STEM courses even before college. For example, Shally shared her experience of how her instructor in a coding course provides more detailed answers when answering men's questions:

*Shally: Sometimes when the girls mess up in the class, he's [the instructor] just like "okay, just share your screen, like I cannot understand what you're talking about" and then when it's the guys, he's like "oh we'll just do this, do this, do this, like..."*

Mila shared a similar experience in her high school chemistry class:

*Mila: When a guy is asking question to him [the instructor] he would be like "oh that's a really interesting question, let's look into it more". For mine, he'd be like "oh it's like we've been over this before" and it's like okay, just because my question is more simplistic doesn't mean that [it is not worth going over] … you know what I mean?*

Mila also mentioned that this high school instructor "*sometimes would be too busy playing a chess game with one of the guys to like actually [bother to even] answer my questions*". We can see from these examples from Amy, Shally and Mila that they felt that the instructors or TAs gave more detailed feedback and answers to male students than they gave to female students, which usually made the female students feel that their questions were not as important or good as male students'. In our interviews, we also found that sometimes female students felt that the instructors overexplained something to women but not to men during office hours, which was perceived by them as being due to the instructors' low expectations of them and thinking that they did not know much physics. This type of behavior can also discourage female students. For example, Lucy shared her experience in physics office hours:

*Lucy: I remember going to office hours with one of my female friends. And we were waiting outside. And then two of the guys that we knew from class came out, and then we went in and like, whenever we'd ask this professor question, he would just kind of like, work out the whole problem for us. Even though we just asked for like one little thing, we had done most of it ourselves. When we asked for like one part of a five-part question, he went over the whole problem from the beginning. And then like we compared notes with the two guys who were there before, and they're like, "yeah, he didn't do that for us".*

Both Lucy and her female friends wondered whether it was because the instructor did not trust their answers to other parts of the question. The fact that the instructor did not answer male students' questions in this way made these female students feel that this instructor had lower expectations of women. In addition, the interviewed female students also reported that instructors sometimes used humiliating sexist language and explicitly showed biases toward women in class. For example, Evelyn shared her experience in a college physics class:

*Evelyn: There's like a group of girls that are also next to each other, they're really good friends. And he [the instructor] called on one of them. And she didn't know the answer. And he was like, "did you read the book?" And she said, "No, I haven't read yet" … And then he was like, "what, so all of you are just in college for the social aspect?" And that was, everyone was just like, kind of like, visibly like, Oh, that was a really weird thing to say to a group of girls who were like friends…*

Evelyn also recalled that because it was only the second day of class, actually no one in the class had read the book yet by adding:

*Evelyn: [the instructor] was like suggesting that maybe like they [the group of female students] are only going to school because they want the image or that they have ulterior motives [for being in physics] or as though they're not really passionate, hardworking scientists, which they absolutely are…I think it's kind of just like ingrained sexism and like, misogyny… I think that there is some part of him that maybe thinks that women don't take it as seriously as men do. And he might not even be aware that he feels that way.*

Evelyn also mentioned that even though there had been times when she noticed that several physics professors' behaviors were not appropriate, she had never said anything to the professors or reported her experiences to superiors because she thought, *"what's it gonna matter?"*. In addition to Evelyn, other interviewed female students noted that they felt hesitant to communicate their feelings to the instructors or report their experiences because they worried that this will influence how the instructors think of them and even jeopardize their course grade. It is clear from the interviews that the lack of safe and effective ways for students to communicate their negative interactions with instructors and TAs and their overall learning experiences is itself very problematic.

### (b) *Letting men dominate the class*

This code refers to situations in which the instructors and/or TAs let men dominate the class and the classroom dynamics dominated by men makes women feel marginalized. Elaine shared her experience in a physics professor's office hours, in which she felt being marginalized in this way.

*Elaine: I wasn't a very big fan of like, how he [the professor] did them [office hours]. Because if you'd like had a question, he gives you like, just like a tiny bit of information and wants you to just know what that means… or like, sometimes you'd ask a question, and he'd be like, aah, this person [a male student in the office hours] can explain that ... that was like, really embarrassing for me… it made me feel like lesser than people who were supposed to be my peers.*

Elaine mentioned that maybe the physics professor thought that having students explain physics to their peers can reinforce their learning, but to her, *"it felt kind of like, degrading, having a [male] peer explain something that I asked to a professor, you know? It felt like, my question wasn't important enough to be answered [by the professor]."* In addition, in the office hours, this professor usually gave *"just a tiny bit of information"* for Elaine's questions, while spending a lot of office hour time answering male students' questions even though many of their questions were not very relevant to that physics course:

*Elaine: I think another problem with the office hours is that there were also a lot of boys, they're trying to show off, and they asked like, very deep questions and stuff that wasn't like really important [for that physics course]… I showed up at office hours to get help with the homework that I needed help with. And they'd be asking totally random questions about*

*like one very, very specific thing. I was just kind of annoyed that they were like talking about something so irrelevant when I needed help with something more tangible. But also, it kind of was annoying that the professor spent so much time answering their irrelevant questions when I had like, actual problems with the material we were learning in class.*

Elaine mentioned that this also happened in her physics TA's office hours and she sometimes ended up wasting a lot of time as in this situation described below:

*Elaine: [The TA's] office hours were like, two hours long. And we spent the first hour answering one guy's question that had nothing to do with what we were learning. So I just sat there for an hour, like trying to interject, but like, my TA, just kept going on and on about this, like, irrelevant subject. And I don't know, I just feel like it's, like, those questions shouldn't be prioritized [in office hours].*

As we can see from Elaine's experiences, the physics offices hours are sometimes dominated by male students. She felt that the professor and TA did not make efforts to balance the time they spent on each student and, on the contrary, they showed more interest in male students' questions, which made female students like her feel marginalized. Elaine noted that after these experiences, her interest in the course decreased and she stopped going to the office hours because *"It just felt like I wasn't supposed to be there. That wasn't the right course for me."*

In addition to office hours, the interviewed female students also reported that they sometimes felt being talked down to or marginalized in group learning situations especially when they were the only female student in the group with several male students. Emma shared an example of her experience as the only woman in a physics learning group with three male students in an online collaboration over Zoom during the pandemic:

*Emma: We had this, like a weekly assignment, and I was the one like writing down the answers and turning in ... I was [the] only one that put my camera on and unmute. [When] I asked questions and then they wouldn't answer them, I would basically have to do it by myself. After two to three weeks, I was like "hey guys I've been writing down the answers for a while, I think it's someone else's turn to do it". So then, as soon as one of the guys started writing down the answers, the rest of the guys started participating…*

Emma further noted that she felt very sad and helpless, and attempted to find resolution by reporting these experiences to her male physics TA:

*Emma: I had like no patience left…it was bad and I started crying and I went to the TA, and I was like "I can't be in this group anymore, they don't listen to me, didn't even call me by my name, they called me 'the girl' to my face … My name is on the [Zoom] screen, it's not that hard" …He [the TA] was just like "It doesn't matter, this is such a low stakes thing, it's just a participation grade. If you don't want to participate with them, just try to practice problems on your own, why are you worried about how your group was doing?".*

As we can see from Emma's experience, the male TA did not help her resolve this problem, and he did not even validate her feelings. Emma further noted that, for the rest of the semester, she just turned off her camera and microphone and worked on her own during the times assigned for group learning, in which students were supposed to co-construct knowledge in a collaborative environment. Emma's example is consistent with prior studies [24,59] showing that in a non-inclusive and inequitable learning environment, female students cannot fully benefit from interactive learning and gender performance gaps can grow.

In addition to being marginalized in physics learning groups, some female students also reported being marginalized in classes dominated by male students. Paola shared an example of her experience in a physics lab course, in which she was the only female student and all the other male students paired with each other and left her to do the experiments alone.

> *Paola: "I was the only girl in that class, which is not ideal…people [the other students] kind of paired off, like, "Hey, do you want to do the photoelectric effect with me?" or "Do you want to do blackbody radiation with me?"… I ended up doing that lab by myself. The first few weeks were fine, because it was like programming and stuff, which I'm obviously comfortable with. And then for the first like, intensive lab that we had to do for that class, I didn't have a partner, so I was doing like the photoelectric experiments by myself, which was just a ton of data, a ton of lab write up. I knew, I can't do this anymore.*

In addition to having to do the lab alone, Paola also mentioned that in this lab class, "*people [male students] don't really listen to you. It's not necessarily intentional. But you suggest something, and people just ignore you…I didn't want to go to that class anymore…I didn't want to deal with the mental stress of that. Or, you know, pay money for a class I didn't want to go to*". Paola ended up withdrawing from the class after several weeks. Even though she told the lab instructor about her experiences before she withdrew, he did not try to improve the learning environment or make efforts to help her.

Michaela's recollections of her physics classes, below, sum up much of what the other interviewed female student discussed in their narratives:

> *Michaela: I never felt like they [her physics instructors] were encouraging women… The men in the class, were kind of like, already dominating conversations, they [instructors] would kind of answer those [men's] questions and go on with whatever they [men] were thinking. They kind of focused on that as opposed to encouraging women to share their ideas and ask questions.*

### 3. Feeling that the physics learning environment is unsafe

In our analysis, 29% of the interviewed female students reported experiences that were coded under the subtheme *feeling that the physics learning environment is unsafe*, which includes two codes: *condescending or intimidating behavior* and *cold calling*.

*(a) Condescending or intimidating behavior*

This code refers to situations in which female students noted that the instructors and/or TAs displayed a feeling of patronizing superiority when answering student questions or caused them to feel fearful about not knowing, which can deter them from asking questions and communicating with instructors and TAs. Madalynn shared her fear of going to office hours for her physics courses:

*Madalynn: I feel like it's intimidating sometimes to go to professor's office hours. I don't want to go in there and look like a complete idiot… Sometimes, in office hours, you get a professor who's like "You're really asking me this question?" It's like, that's just what I want to avoid.*

Angelina shared an experience about feeling intimidated in a synchronous online class.

*Angelina: The teacher would be harsh when people asked "not good" questions… He'd be like "You don't understand what you're asking!", then he might call you out for it, so quite a few people were like a little scared to ask questions, because if you don't ask the right question, you might get him annoyed.*

As we can see from Madalynn and Angelina's experiences, they were not only discouraged but also intimidated by their instructors. Angelina added that the instructor disabled the chat function, so students could not type their questions in chat. In addition to implicitly being harsh on students, the way instructors ask or answer questions can also be intimidating to students. For example, Mary shared how intimidated she was when she saw how her physics instructor answers students' questions in his office hours:

*Mary: He [the professor] had been talking about like, you should answer questions until you don't know…and then you should look for ways to figure out how. If you want, I can help you with that...And while I was there [in his office hours], this other kid in the class had come. [The professor] would basically ask him questions until he didn't know stuff… And I was like, intimidated by this other kid in there doing that, well, I had a really dumb question to ask...*

Even though Mary's instructor may want to use this Socratic method to encourage students to find out the answers by themselves, the way he did it during office hours might make students even more nervous about not knowing and feel unsafe to expose their struggles. This is particularly true for students from the marginalized groups in physics such as women.

In addition to feeling intimidated by instructors, the interviewed female students also reported that sometimes physics instructors may create a stressful learning environment in which top performance is over emphasized. For example, Evelyn shared an experience in one of her physics courses.

*Evelyn: He [the professor] would announce names and grades of high scorers [in front of the class], which I thought was kind of weird because it was like if you weren't in the top and then everyone knew what your score was almost… I don't know what the purpose of that was, maybe like to chase the glory of being, having your name called in class, which is just kind of weird to me…I think this is like a weird environment… [my score] shouldn't be anyone's business…I think everyone [other women she talks to] thought it was really weird.*

Evelyn further mentioned that announcing names of top students may promote an unhealthy competition. This may foster a fixed mindset in the classroom, especially when most of the top students are men, because due to societal stereotypes and biases, the words "brilliance" and "intelligence" are associated with men.

In addition, the interviewed female students also reported that some of their instructors were condescending when answering questions. Hailey shared her experience in her electricity and magnetism course:

*Hailey: I just did not feel comfortable with him [the instructor] at all. I didn't feel like I could ask him questions without fear of like being mocked or something like that. So, if I had any questions about that, I would talk to my friends… this professor [also] didn't mind mocking like humanities majors and classes and stuff like that, and saying like, "Oh, like they're not as hard as physics and stuff" …he could be pretty condescending even in class. I just didn't really want to be around it.*

Hailey noted that she did not think that talking to this professor would have helped her because, *"he wouldn't have listened…there just seemed to be no connection between him and the students… it didn't really seem like he had any concern about making a positive learning environment"*. She mentioned that it is also true for many other physics instructors because *"They seem like set in their ways, and not really open to other opinions"*. Hailey's experiences indicate that it is necessary to provide professional development and train physics instructors to help them realize their responsibility in building an inclusive and equitable learning environment and they need to make intentional efforts to build such an environment in which all students feel safe and can thrive.

*(b) Cold calling*

This code refers to situations in which women described instructors calling students to answer questions at random or asked them to share answers in front of their peers. Even though the instructors may think this is a good strategy to engage students, many interviewed female students noted that they did not like it and this approach made for a very stressful environment for them. For example, Evelyn shared how students like her felt when a physics instructor randomly called on students.

*Evelyn: He [the instructor] had this Socratic method, which is like he kind of puts people on the spot. It is kind of stressful, because it feels like going in class… you're just like a sitting duck, waiting to be called out…I know a lot of people don't like that and think it's like a stressful environment.*

Maya shared a similar experience in which she and her female peers were concerned about being called randomly in a physics course:

*Maya: I've spoken with all the women who I sit with at the beginning of the semester. When we heard that he [the instructor] likes to call on people at random, there was one spot in like an initial thing where you wrote your name, year, background and stuff. And then [there was a question] "Do you have any concerns about this course?" All of us were like,*

*can we write that we don't want to be called on at random? We literally talked about writing that in the concerns for the course.*

Maya mentioned that despite writing that, they still ended up being called randomly to answer questions in the course. She further noted that the reason they do not like cold calling is because they worry about getting the answers wrong in front of the whole class adding, *"but I guarantee our professor isn't even really thinking about it… They're just doing it"*.

In addition to being randomly called in class, similar situations can arise in office hours. For example, Mary shared her experience in a physics instructor's office hours:

*Mary: His office hours were just largely very overwhelming to attend. He would ask you to write stuff on the board in front of the whole group. It was very intimidating, especially because I was there to ask for help. And so like, I didn't know how to do the stuff he was trying to ask us to do.*

Mary further noted that male peers were already showing off and she did not want her male peers to think she is not smart and *"I felt like I had to prove it to them that I was smart enough"*. As we can see from the interviewed women's experiences, cold calling may increase students' anxiety, particularly if they are from marginalized groups, which is consistent with prior studies [60,61]. Moreover, the anxiety caused by cold calling may increase stereotype threats for students from underrepresented groups such as women.

### 4. *Feeling negatively recognized about their abilities and potential*

In our analysis, 16% of the interviewed female students reported experiences that we coded under the subtheme *negative recognition for students' ability and potential*, which includes two codes: *underestimating students' ability* and *fixed mindset about students' potential* [62].

*(a) Underestimating students' abilities*

This code refers to situations in which instructors doubt students' ability to do well in a task or have low expectations of students. For example, Paola shared an experience of consulting with her physics department academic course advisor about the courses she wanted to take.

*Paola: I said [to the advisor], "I'm gonna take that computational physics class… it's like coding and I have taken the entire CS [computer science] core [courses], so that's fine. And I've programmed in Python for work". And he's like, "Well, it's really project heavy." … It just felt like he's just trying to dissuade me. He just kind of kept saying things like, well, it's really project heavy. And I was like, I'm taking CS 1501, which is like an algorithm's implementation course, which is like a weeder class for CS. I got a 90 for my midterm. It's really project heavy. I can do coding projects like I've been doing that for a while. I've had to code like three years ago, it's fine… He's like, "well, it's gonna be hard" and I was like, what? I know how to code. You can look at my transcript, I've taken more programming classes than probably most of the guys. So why are you saying this to me?*

As we can see from Paola's experience, the academic course advisor implied that the course she wanted to take was too difficult for her. Even though after Paola showed that she had enough coding experience to take this course, the advisor still doubted her ability, which made her frustrated. Paola further added that this actually was not the first time she had such an experience pertaining to physics related course work. Her previous academic course advisor in the first year before she declared the physis major (students usually declare physics major in their second year) also didn't trust that she could take an advanced calculus course, "*He was like, are you sure you should take calc III?*". Even though Paola took this course anyway, she said that she was *"annoyed"* by people who keep doubting her ability.

During the interviews, other female students also related similar experiences and they also reported examples of other female students going through similar experiences. For example, Evelyn shared an experience of one of her female peers who did undergraduate research that required extensive coding being dissuaded from taking the computational physics course with another core class in physics.

> *Evelyn: My friend went to her advising appointment. She wanted to take the computational methods [in physics] that [is] like coding class, and her advisor was like telling her she's really going to struggle with it, like it's really difficult and like he doesn't think that she should take it with another core class… And she was really offended because…she was like, "why does he think that I'm not gonna be able to handle this coding class". And it's particularly strange because her [undergraduate] research is coding. She codes in Python every day for hours, and she's really, really good at it. So it was weird that he [the advisor] would kind of assume that she's gonna really struggle with this class and she couldn't handle the workload with other things. Particularly because other people like men were like, "he's [the advisor] never said anything about that to me" …that was really frustrating."*

Evelyn mentioned that witnessing this makes her think that the professor may have lower expectations of female students and added that "*Actually, it happened two semesters in a row. He did it again this semester. He was like, I think you should wait on that class, because coding is very difficult. And she's in it now and it's a breeze.*" We note that the experience of Evelyn's friend is similar to Paola's. However, since Evelyn is a senior and Paola is a second-year student, we believe that these two episodes are more likely to be about different women even though there is no way to confirm.

Many interviewed female students mentioned that the experiences of being underestimated make them really want to prove to others that they deserve to be in physics, and they are as capable as men are, but this feeling sometimes caused extra pressure on them as female students in physics courses. These findings are consistent with prior studies showing that negative stereotypes about women in physics may cause women to assume that they have to make extra efforts to succeed in physics relative to male students and their achievements are not a reflection of how good they are in physics unlike the achievements of "successful" men who excel in these fields without making effort [58]. Likewise, women in physics courses might undergo additional stress and struggle to demonstrate their skills to be valued equally as men in a classroom in which they are underrepresented. These pressures may partially explain the finding in a prior study that female students are more likely to drop out of STEM majors such as physics than their male peers who earn the same grades [63].

In addition to lack of recognition and encouragement, the interviewed female students reported that sometimes micromanagement can also make them feel that physics professors do not trust their ability. For example, Mckinley shared her experience working as an undergraduate researcher with a physics professor in his lab.

> *Mckinley: If I was asking him [the professor] a question on how to input a command into a Linux computer… He was like leaning over my shoulder and like telling me one key at a time what to type instead of just trusting me to be able to spell the word … It was weird being micromanaged in the places where I've felt pretty confident in myself. So it was just like a frustrating experience.*

Mckinley noted that on one hand this professor micromanaged her on the things that she was confident with, on the other hand, there was no guidance when she really needed help with the research. She felt frustrated and did not think that she was given a project that played to her strengths.

### (b) *Fixed mindset about students' potential*

This code refers to situations in which female students described instructors emphasizing brilliance rather than effort or thinking students have a set amount of ability. For example, McKinley reported that some of her physics instructors think that only some people can do well in physics.

> *Mckinley: And then there really are professors here who think that some people will just never be able to learn certain concepts.*

Having a fixed mindset about students' ability can be especially harmful for students from marginalized group (such as women in physics). When instructors convey the message that physics is not for everyone, due to the pervasive gender stereotype in physics, women are more likely to think that they are the people who do not have the innate ability to excel in physics. For example, Paola described her fear about being judged by her instructors for being a woman.

> *Paola: [Some male physics professors] tend to have an air where it's like, this is a naive solution, this is so trivial... Sometimes there's condescension, which makes you not want to ask questions anymore… It can be especially hard when you have that question in the back of your mind which is like, are they being condescending because they think I can't do it, or because I'm a woman?*

In our interviews, Madalynn shared the response of her high school physics teacher when she told him that she is pursuing a physics degree, which may also be due to his fixed mindset.

> *Madalynn: I went back a few times. And every time he [her physics teacher] was like, surprised that I did a physics degree. I don't really know why … He's always kind of like, oh, really?*

Prior studies have shown that instructors' mindset about whether all or only some of their students can excel in their courses can influence students' motivation and achievement, and underrepresented students are more likely to be demotivated and have negative experiences in classes taught by fixed mindset instructors [62].

## B. Impact of negative perceived recognition or lack of positive recognition from instructors/TAs on students' self-efficacy and interest

In the last section, we discuss different types of negative perceived recognition or lack of positive recognition from instructors and/or TAs. In our interviews, we found that these experiences of not being appropriately recognized often had a negative impact on female students' learning and their physics self-efficacy and interest. In addition, the lack of appropriate recognition can influence female students' persistence and retention in physics and other STEM majors.

### *1. Impact on students' self-efficacy in physics*

In our interviews, we found that female students' self-efficacy is often negatively influenced when they feel that their questions or struggles are belittled by instructors or TAs. For example, Raina shared the feeling students might have when physics instructors assume that something is easy for students:

*Raina: The professors just expect you to completely understand it and so that can be very discouraging sometimes because, they'll just like go on and assume you know things…That can cause a lot of students to just be like I'm doing so poorly, like I am not confident and stuff like that.*

Shally shared a similar experience in which her physics professor assumed the new physics knowledge is easy for students and did not give enough time for students to do the classwork:

*Shally: Sometimes when he [the professor] explains some of the new stuff that we're learning, he would be like "oh yeah this is easy, you should be able to get this in like five minutes"… He kind of has this demeanor that it's so easy and you should be able to get it so quick and if you don't then you're not good.*

In addition, feeling belittled for struggles or questions can also cause students to doubt their ability to do well in physics. For example, Hailey shared her feelings when she received negative recognition from her physics instructor.

*Hailey: [I am] like having this professor who didn't seem [to think] my questions were valid…It just made me feel like stupid… and like other people in the class were smarter than me and like I wasn't capable of doing well.*

Evelyn had similar feelings when she was told by physics instructors that something she had struggled with was easy:

*Evelyn: They [physics instructors] say like, this is trivial, this is easy, this should be obvious. So then, if you don't think that, you immediately, you're like, oh, there's something wrong with me. I'm like, I'm missing this super obvious thing.*

As we can see from these examples, when physics instructors belittle students' difficulties, questions or struggles, students' self-efficacy can be negatively impacted, and they may even think "*there's something wrong with me*". The situation can be worse if the same instructors show more interest in and patience for men's questions because it can convey to women that they are not as capable as men and make female students feel marginalized. Amy shared her feelings after she was consistently ignored by her male TA, who *"brushed off"* her questions and answered the questions of male students in her group.

*Amy: …anytime that I made a small mistake, I felt like it was completely larger than it was, and I felt like I had made a grave mistake and I was not as smart as my classmates. Even if it was just typing something incorrectly into my calculator. Every little thing that I did incorrectly or every time that I wasn't listened to, I thought it was because of me, and because of my academic success and my personal knowledge…*

As shown in the examples above, feeling belittled for questions or struggles and feeling marginalized due to differential gender dynamics are two major types of negative perceived recognition that impacted female students' self-efficacy.

## 2. Impact on students' interest in physics

In addition to the impact on students' self-efficacy, our interviews show that negative perceived recognition or lack of positive recognition from instructors and/or TAs can also negatively influence students' interest in physics. For example, Amy noted that her experience of being ignored by the male TA who paid more attention to male students' questions made her feel that she did not belong in physics and also made her start to lose her interest in physics and her desire to major in engineering because physics courses are required for engineering students:

*Amy: [The TA] is someone who is supposed to be a role model for me, and it's supposed to be an outlet for me to get help where I don't necessarily feel comfortable getting help…They made me feel as if I didn't belong, which made me question, why am I here in the first place? Am I really interested in this? Do I really want to be here, meaning like in the engineering school…? And for me that [the experiences in the physics course] was a big reason why I was thinking of switching out of engineering…*

As we can see, being treated differently by physics instructors or TAs as compared to male students can negatively impact female student's sense of belonging and interest, which may further influence their persistence and retention in physics and other related STEM majors. In addition to being treated differently by instructors or TAs as compared to male students, we find that female students' interest in physics can also be negatively impacted if instructors let the class be dominated by men. For example, Katie shared her experience in a physics course.

> *Katie: I guess in that class there were a lot of people who would just ask questions to show that they're smart like that they know what they're talking about. I was like, I don't even understand your question, I don't know what you're asking ... They were all male, I don't even know how many girls were in that class, but it seemed like you only heard from the boys ever. It just kind of seems like they're [instructor and male students] like off in their own little world talking about something that the rest of us don't understand.*

Katie further added that the physics instructor talked very fast in class, and she had to re-watch the lecture videos and put it on half speed (these were online classes during the pandemic). At the same time, the way some of the male students who dominated the class talked made her feel that the male students in general seemed to know what was going on in the class. Moreover, the professor gave very low scores on exams. Katie said that she put in a lot of effort in this course but still got 30% on a test and 60% after curve. These experiences strongly impact her interest in physics.

> *Katie: I hated physics. I just hated it so much. Anytime somebody would bring it up, I would like shudder… I cannot put into words how much I hated that class. I was like I can't do engineering, because I don't like this class ... I really was like about to switch and do something else…I was like maybe this isn't for me, some people really seem to know what they're talking about, like maybe it's just a me problem.*

As we can see, these experiences not only influenced Katie's interest in physics but also her self-efficacy since it made Katie think that it is her problem not being able to understand the discussion between the male students and the instructor. In addition to students' experience in physics courses, we found that not being recognized appropriately in research lab can also negatively impact students' interest in the research field. For example, Mckinley shared her experience working as an undergraduate researcher with a physics professor in his particle physics lab:

> *Mckinley: He would just give me like a list of a bunch of stuff to do and never tell me when he wanted it done. And then a lot of the time, by the time I had done some of them, he wouldn't care about it anymore.*

In addition to feeling that her work was not recognized by the prefessor, Mckinely also mentioned that she did feel safe to ask the professor for help in her research.

> *Mckinley: I felt very, very intimidated by asking him for help ... He [the professor] was really careful to be like, everybody can learn physics, but I feel like he didn't actually put that into practice in his own lab, which was frustrating… I learned a lot from that experience, but I'm definitely glad that it's over. I don't really think particle physics is the thing I want to study.*

Finally, Mckinley left that lab since she thought she was not interested in particle physics anymore. As we can see in the above examples, feeling marginalized due to differential gender dynamics, feeling belittled for efforts, and feeling the learning environment is unsafe can negatively influence female students' interest in physics.

### *3. Impact on students' persistence in physics*

As mentioned earlier, Amy and Katie considered switching out from an engineering major because of their negative experiences in physics courses. Our interviews suggest that physics majors may also switch out of physics due to negative recognition. For example, Elaine shared many experiences pertaining to how she felt negatively recognized in physics courses, for example, how her questions were belittled by physics instructors in office hours and how the physics courses were dominated by men. She mentioned that the only reason she is still a physics major is that she has dreamed of becoming a physicist from age 12 and she wrote her dream down, keeps it with her and reads it whenever she feels discouraged and wants to quit.

*Elaine: I think what I'm trying to say is, I would not have been here still if I hadn't been so focused on my reasons for being here. Because there have been so many times where I've been like, I don't even know why I'm doing this…*

Elaine also mentioned that she had female peers switching out of the physics major because of lack of recognition and support.

*Elaine: I know four separate girls who were planning on majoring in physics and astronomy, but after the intro courses, they were too discouraged and switched out… which is really sad because I think they could have done it…it's hard, since there's not a lot of support.*

As we can see, negative recognition from instructors/TAs can impact students' self-efficacy and interest in physics, which have been shown to be very important for student engagement, performance, and retention in physics. Our interviews show that female students may switch out from their majors because of the discouraging experiences they had in physics courses. Therefore, instructors and TAs should realize that it is their responsibility to appropriately recognize and support students and make intentional efforts to build an inclusive and equitable learning environment in which all students can thrive.

### C. Positive recognition from instructors/TAs and its influence on students' self-efficacy

In previous sections, we discussed different types of negative perceived recognition/lack of positive recognition and their impact on female students' self-efficacy and interest. In our interviews, when explicitly asked about positive recognition from instructors and TAs, only four interviewed female students shared their experiences of being positively recognized by physics instructors, which helped their learning and boosted their self-efficacy. In addition, some female students also shared their experiences in other courses (not physics) that made them feel positively recognized. In this section, we will discuss examples of how some instructors positively recognized students.

In our interviews, we found that explicitly recognizing and validating students' ability can help them build their physics self-efficacy. For example, Hailey shared a positive experience communicating with her quantum mechanics instructor:

*Hailey: My quantum professor, he's doing an excellent job at seeing what the students want…I feel like really supported in that environment, especially like, when I go to office hours… this professor literally said to me, like, "I think that you're just struggling with some of complex math stuff, but you're good on the concepts, you're doing well for where you should be". That was really encouraging because quantum is difficult for me at this time and it would definitely be easy to just feel overwhelmed…But the fact that this professor literally was like, "No, you are doing well, and you're capable of doing this.", it just makes me want to try so much more. And it makes me feel like I really am capable of doing it.*

Hailey further added that this kind of validation is not typical of what she is used to getting from physics instructors so it feels strange to her:

*Hailey: … I feel like supported in the class, [it] is just like so strange, but in the best way…I enjoy the class so much more and I feel really motivated to study especially compared to my other core physics classes…*

As we can see, positive recognition from instructors has the power to encourage and motivate students to work hard and boost their self-efficacy. Hailey used the word "strange" to describe her feeling of being supported by only one of her physics course instructors, i.e., such affirmation was not typical which is sad because this should be the situation in all of her physics courses.

In addition to explicitly recognizing students' ability, encouraging students to pursue their goal is also positively recognizing them. For example, Katie shared her experience of talking to a physics instructor when she felt overwhelmed:

*Kaylah: I really liked him as a professor…I was like, have a hard time getting concepts and stuff and I went to his office hours a couple of times at the end of the semester. I was like telling him how much I was kind of struggling or like worried about this class. And I told him, I wanted to be a physics and astronomy major. And he was like, "Oh, I don't think you should be discouraged by this one class. And if you really want to do this, you should keep trying". And so I [kept trying and] did pass that class…*

As we discussed in earlier sections, interviewed female students often reported that asking questions to some physics instructors could be quite intimidating. On the other hand, Evelyn's account below suggests that a physics instructor who acknowledges that struggles are normal may reduce students' fear of asking questions or making mistakes:

Evelyn: *When I was in his class, he told me he went to Harvard I think and he was like, "Yeah, I got like a 20% on my first test"… that just shocked me cuz I, I was like, wow… this person that I thought was just like this, you know, perfect academic like genius also struggled in the same way that I struggled and is now like a professor, you know, doing research… that just really altered my perspective of it. Because now it's easy to think of your professors as just like a flawless genius and they're just like constantly judging and grading you, but it's helpful when they open up and tell you what their experiences are and that your experience is normal, and you don't need to freak out about whatever you're thinking…*

In addition to physics instructors, the interviewed female students also shared experiences of feeling positively recognized by instructors of other courses. For example, Suzie shared her experience in her chemistry class:

*Suzie: I like to hear when my professor knows what students tend to struggle with and why they struggle with something. For example, if she sees something that in general students have trouble with, she'll say, students tend to have trouble with this part specifically, so focus more on this when you're studying. And I enjoyed that because it tells me specifically where I might struggle with and I don't feel as bad when I do struggle with that topic, because I know that other students in the past tend to struggle with it...*

As we can see, Evelyn's and Suzie's instructors help students to realize struggles are normal by sharing their own experience or sharing prior students' struggles. Sometimes, instructors can even help students by normalizing struggle by showing in a visual manner that many students have similar questions that they do not know the answers to:

*Raina: My chemistry teacher would always encourage us to ask questions…Like one time, when someone had a question and asked it, he stopped the class and was like, "How many of you are also thinking this question?" and like half the class raise their hands. So it was like half of the class had the same question, but everyone thought that they were like the only person that had the question, so it was like very helpful to like visually see how so many other people were also confused on the same topic like. I feel like that definitely made it very easy for me to just like realize, oh like I'm not the only one that's confused all the time.*

These examples show that after students realize that struggles are normal, they are not the only ones who are confused, and the difficulties can finally be overcome by working hard, they tend to be less nervous about not knowing or making mistakes. This can be helpful for developing their self-efficacy and interest. In addition, when students ask questions, instructors who recognize students' effort and ideas come across as more approachable for students. For example, Mila shared how her math instructor explicitly recognized students' questions and/or thoughts:

*Mila: …my math professor, he always says, "you know, that's a really good question let's look into it and kind of go through it" rather than just, "this is the answer"…And then he kind of makes you feel like, "Okay, thank you for thinking about this in a different way, we can answer this so you can understand it better", not just I'm answering your question to shut you up and get you out of the way.*

Helen shared a similar positive experience in her math methods in physics class, which influenced her self-efficacy in this class:

*Helen: The [math methods] professor is very open to questions, and he's really nice … he will not make you feel like stupid for asking the question or anything like that. And like, anytime you want to, like go up to his office and ask a question, he is happy to help. And that just like makes a difference in how well I feel like I can do in the class.*

The interviewed female students also mentioned that interpersonal communication with instructors and advisors is very helpful for their learning because their personal life and academic life are so deeply intertwined. For example, Charlize shared her experience with her academic advisor through her sorority.

*Charlize: She's really nice about like, sitting down, just because I had some classes that didn't go so well. And so she sat down with me… just to like plan out the rest of my steps to graduation and things like that… Every time we go, we print out another copy of the physics major requirements and figure out what I've done and what I haven't done and how to make my next semester, like, minimize the stress from classes and things like that.*

Charlize added that personal attention on specific issues she is having has been extremely beneficial to her:

*Charlize: That's been really helpful just because I feel like it's like, for me, it's really having her because she asks questions about my life and like I can tell her all like the issues that I'm having and things like that. And she'll listen and like talk to me and like, more like interpersonal interaction added to that has been a really helpful thing for me…*

As we can see from Charlize's experience, in addition to clearly communicating that difficulties are a normal part of learning and being open to questions, it is also important for advisors and instructors to provide students with appropriate guidance and support and let them know that they believe that they can do well, and they are always there to support them and brainstorm with them about the challenges they may be facing.

## V.    SUMMARY AND DISCUSSION

In this study, we conducted interviews with female students enrolled in physics courses to investigate their experiences and perceived recognition from instructors and TAs and the relationship between their recognition and the other two dimensions of physics identity, i.e., self-efficacy and interest. In particular, with regard to the first research question focusing on female students' perceptions of different types of recognition from instructors/TAs in physics courses, our results indicate pervasive existence of negative perceived recognition or lack of positive recognition from instructors/TAs in the current learning environment. In particular, the negative perceived recognition or lack of positive recognition was categorized into four categories: (1) feeling belittled for questions or efforts, (2) feeling marginalized due to differential gender dynamics, (3) feeling that the physics learning environment is unsafe, and (4) feeling negatively recognized about their abilities and potential. We find that the most common negative perceived recognition from instructors and TAs is feeling belittled for questions or efforts. When students are confused about a new concept or have questions, the words "easy", "obvious" or "trivial" from instructors and TAs can convey to students that if they cannot do such easy problems on their own, they may not be smart enough to do physics. This situation is exacerbated when the same instructors/TAs show interest in other students' questions (especially those from male students) particularly because due to societal stereotypes and biases, the words "genius" and "brilliant" are associated with men, and thus the dichotomy in the instructor/TA responses to female and male students' questions can reinforce the negative perceived recognition by women [47]. Moreover, our interviews show that instructors' use of the words "easy", "obvious" or "trivial" for physics problems can be learned quickly and used abundantly by students (particularly male students) when communicating with other students and contributes to the toxic culture of physics.

Our second research question focused on how female students' perceived recognition from instructors/TAs shape their self-efficacy and interest. According to our interviews, feeling belittled for questions or struggles and feeling marginalized due to differential gender dynamics can negatively impact female students' self-efficacy. For example, we find that if instructors and TAs convey a doubt about female students' ability to do well in physics or convey a low expectation of them, students can also start doubting their ability even more. These findings are consistent with prior quantitative studies showing that women with A grades have the same self-efficacy level as men with C grades [43]. In addition, negative perceived recognition or lack of positive recognition from physics instructors or TAs can also impact female students' interest in physics. In particular, our interviews show that feeling marginalized due to differential gender dynamics, feeling belittled for efforts, and feeling the learning environment is unsafe can negatively influence female students' interest in physics or in a physics field. For example, we find that if instructors let the class be dominated by men and let women feel marginalized, female students are less likely to enjoy working with others and engage in learning. As a result, these female students explicitly mentioned that these experiences impacted their sense of belonging and interest in physics. Moreover, our interviews show that the experience of not being appropriately recognized by instructors and/or TAs may also influence female students' retention and persistence in physics and other STEM majors. We note that perceived recognition, self-efficacy and interest are three dimensions of physics identity, which has been shown to influence students' career decisions as well as short- and long-term academic goals [2,5,27].

We note that even though instructors and TAs sometimes negatively recognize or fail to recognize or validate students unconsciously, they still impact students' self-efficacy and interest. Therefore, it is important for instructors or TAs to realize that it is the impact on the students that matters and not their intention. Eileen Pollack, the first woman to get a BS degree in physics at Yale University, has provided an excellent illustration of the effects of a lack of positive recognition in her memoir [64]. As a child, Pollack wanted to be a theoretical physicist, but after graduating, she eschewed her childhood dreams and decided instead to pursue graduate work in English. In her book, she recounts how she felt when she was dismissed by her instructors and even her undergraduate thesis adviser after solving a theoretical problem for her thesis: "When at last I found the answer, I knocked triumphantly at my adviser's door. Yet I don't remember him praising me in any way. I was dying to ask if my ability to solve the problem meant that I was good enough to make it as a theoretical physicist. But I knew that if I needed to ask, I wasn't." There are many Eileen Pollocks out there who will not narrate the impact of the negative recognition or lack of recognition on their career trajectories in memoirs. Therefore, it is important for instructors and TAs to make intentional efforts to appropriately recognize and validate students.

## VI. IMPLICATIONS

Our study discussed four types of common negative perceived recognition or lack of positive recognition by female students from instructors and TAs, which can help physics educators to be intentional about not negatively recognizing students and focus on positively recognizing and validating students, which is particularly critical for students from marginalized groups such as women. Our findings can also help researchers to further understand the role played by perceived recognition in shaping students' self-efficacy and interest.

In our interviews, only four interviewed female students shared experiences of being positively recognized by even a single physics instructor. Some other female students shared their experiences in other courses (not physics) that made them feel positively recognized. These experiences include, for example, situations in which instructors/TAs were open to questions and positive interpersonal communication with students, acknowledged that difficulties are normal and surmountable, and were always there to support students in their learning. These findings can serve as inspiration for physics instructors to make intentional efforts to positively recognize their students. A prior study found that the synergy between explicitly and implicitly recognizing strategies used by instructors is a critical feature of positive recognition that can be internalized effectively by the students [65]. For example, instructors can explicitly recognize students by directly acknowledging their efforts and questions and expressing faith in their ability to excel. They can also implicitly recognize students by valuing their opinions and assigning a leadership position or a challenging task to students in small groups that makes them feel excited [65]. In addition to positive recognition, instructors should be careful not to give unintended messages to students, e.g., by praising some students for brilliance or intelligence as opposed to their effort since it can convey to other students that they do not have what is required to excel in physics [34]. We emphasize again that in any professional development workshops for instructors/TAs focusing on these issues, it is important for instructors/TAs to reflect upon and internalize that it is not their intentions that matter but the impact they are having on their students.

During our interviews, female students noted that the physics courses and office hours were usually dominated by men, and some even reported being talked down by their male peers. Also, in most cases, instructors and TAs did not intervene to correct students' behaviors and support female students, and sometimes instructors and TAs belittled the female students by showing no interest in answering the questions that these students asked them. As a result, these female students explicitly mentioned feeling negatively recognized by instructors/TAs in physics courses. A recent study [66] shows that in an equitable and inclusive physics department, when students fail to interact with each other equitably, faculty members intervene in the student-student interactions and insist on students' behaving appropriately and learning the norms. This type of intervention by instructors and TAs in which they take responsibility to protect students from, e.g., sexist and racist microaggressions has the potential to create a more welcoming environment and ensure that all students, regardless of the demographic groups they belong to, can excel.

In addition, in our interviews, we found that cold calling in physics courses can also make female students feel unsafe. When instructors or TAs put female students on the spot, they often feel very nervous and do not want to make mistakes in front of their male peers, which can cause anxiety and rob students of their cognitive resources (since working memory where information is processed while problem solving is limited and part of that working memory can be taken up by the anxiety) and influence students' performance [24,67]. A recent study introduced a novel classroom participation approach called the warm call [68], in which instructors invite students in advance to report out from a think-pair-share or group activity, help students review their prepared response, and provide opportunities to opt out. This method may help reduce the pressure on students while still providing opportunities for all students to share their thoughts. It should be noted that the authors of the warm-call approach emphasized that for this method to be successful, instructors need to ensure that the classroom environment is inclusive and welcoming. If the students do not feel safe or feel that their voices and contributions are valued, they will likely be disinclined to participate in this optional process.

Our qualitative study suggests that being reluctant to answer questions students ask assuming something should be easy for them or micromanaging by providing extremely detailed response when they ask very specific questions (conveying to students that they do not even know any background materials related to it) can both make marginalized students such as women in physics feel negatively recognized. Prior studies suggest that scaffolding support (i.e., appropriate feedback and support provided promptly as needed) is most effective when the support is matched to the needs of the learner [69-71]. Therefore, it is important that instructors/TAs know students' current level of knowledge and provide appropriate support. For example, when a student has specific questions, instead of immediately providing response, the instructor could first ask the student to show their work and reasoning process thus far as well as what they may have been planning to do next and what they were unsure about. Then, instructor can provide help based on students' current level of knowledge rather than making students feel micromanaged and underestimated. In addition, when helping students, instead of working out very detailed solutions for them, the instructor could provide hints or prompts to stimulate students' thinking and encourage them to jointly construct the solution with them, which can not only enhance the effectiveness of learning but also provide students with a greater sense of accomplishment [72-74].

In addition to providing appropriate support, instructors having and conveying a high expectation of all students is also very important [75,76]. Prior studies have shown that students tend to internalize the beliefs teachers have about their ability [77,78]. Instructors should communicate their expectations with students and express their belief about all students' ability to achieve the expectations by working hard and working smart as well as taking advantage of all of the resources. It should be noted that having a high expectation of students does not mean instructors should assume that students know everything and will not have difficulties. On the contrary, instructors should recognize students' difficulties and also help students understand that difficulties are normal and are opportunities to improve rather than a sign of lack of ability [79]. Moreover, instructors should let students know that their instructors will always be there to help and support them and display it in their actions.

## ACKNOWLEDGMENTS

This work was supported by Grant No. PHY-1806691 from the National Science Foundation. We would like to thank all students in physics courses who interviewed with us and Dr. Robert P. Devaty for his constructive feedback on the manuscript.